\documentclass[twocolumn,superscriptaddress]{revtex4-1}
\usepackage{lipsum}

\usepackage{dcolumn}% Align table columns on decimal point
\usepackage{bm}% bold math
\usepackage{color}

\def\ltape{\hbox{\ $<$\hskip -8pt\raise -4pt\hbox{$\sim$}\ }}
\def\gtape{\hbox{\ $>$\hskip -8pt\raise -4pt\hbox{$\sim$}\ }}

   \ifx\pdfoutput\undefined
   \usepackage[dvips]{graphicx}
   \else
   \usepackage[pdftex]{graphicx}
   \fi
   \usepackage{epstopdf}
   \DeclareGraphicsExtensions{.pdf,.gif,.jpg,.eps,.png}
  
   \usepackage{amsmath}
  \usepackage{amssymb}
\pdfoutput=1

\begin{document}
%\linenumbers

\title{First-Principles Theory of the Rate of Magnetic Reconnection in Magnetospheric and Solar Plasmas}

\author{Yi-Hsin~Liu}
\affiliation{Department of Physics and Astronomy, Dartmouth College, Hanover, NH 03750, USA}
\author{Paul~Cassak}
\affiliation{Department of Physics and Astronomy, West Virginia University, Morgantown, WV 26506, USA}
\author{Xiaocan~Li}
\affiliation{Department of Physics and Astronomy, Dartmouth College, Hanover, NH 03750, USA}
\author{Michael~Hesse}
\affiliation{NASA's Ames Research Center, Moffett Field, CA 94035, USA}
\author{Shan-Chang~Lin}
\affiliation{Department of Physics and Astronomy, Dartmouth College, Hanover, NH 03750, USA}
\author{Kevin~Genestreti}
\affiliation{Southwest Research Institute, Durham, NH 03824, USA}

\date{\today}
\begin{abstract}
{{\bf Abstract:} The rate of magnetic reconnection is of the utmost importance in a variety of processes because it controls, for example, the rate energy is released in solar flares, the speed of the Dungey convection cycle in Earth's magnetosphere, and the energy release rate in harmful geomagnetic substorms. It is known from numerical simulations and satellite observations that the rate is approximately 0.1 in normalized units, but despite years of effort, a full theoretical prediction has not been obtained. Here, we present a first-principles theory for the reconnection rate in non-relativistic electron-ion collisionless plasmas, and show that the same prediction explains why Sweet-Parker reconnection is considerably slower. The key consideration of this analysis is the pressure at the reconnection site (i.e., the x-line). We show that the Hall electromagnetic fields in antiparallel reconnection cause an energy void, equivalently a pressure depletion, at the x-line, so the reconnection exhaust opens out, enabling the fast rate of 0.1. If the energy can reach the x-line to replenish the pressure, the exhaust does not open out. In addition to heliospheric applications, these results are expected to impact reconnection studies in planetary magnetospheres, magnetically confined fusion devices, and astrophysical plasmas.}

\end{abstract}


\maketitle

\section{Introduction}
Magnetic reconnection converts magnetic energy into plasma thermal and kinetic energy in laboratory, space and astrophysical plasmas.
Two major and largely separate endeavors have been pursued to quantitatively predict how the plasma is energized by reconnection \citep{wygant05a,drake06a,birn10b,aunai11a,egedal12a,eastwood13a,eastwood20a,shay14a,bessho14a,dahlin14a,FGuo14a,shuster15a,yamada15a,SWang18a,FGuo20a,XLi21b} and the rate at which reconnection proceeds \citep{sweet58a,parker57a,petschek64a,shay98a,birn01a,rogers01a,bessho05a,cassak07a,hesse11a,yhliu14a, tenbarge14a,stanier15a,stanier15c,comisso16a,yhliu17a,cassak17a,yhliu18a,TKMNakamura18a,genestreti18b, RNakamura18a, KHuang20a,XLi21a}.
Nevertheless, the linkage between these two fundamental aspects of reconnection is missing.
%In this work, we show there is a linkage between these two fundamental aspects of reconnection, which allows us to calculate the collisionless reconnection rate from first-principles. 
The most critical question in understanding fast reconnection is what localizes the diffusion region (DR) \citep{Biskamp2001} (i.e., what makes it far shorter than the system size), giving rise to an open geometry of reconnection outflow. Petschek's model \citep{petschek64a} provides a valid steady-state solution for such an open outflow geometry, but it fails to provide a valid localization mechanism in the uniform resistivity magnetohydrodynamics (MHD) model \citep{biskamp86a,uzdensky00a}; reconnection in such system always results in a system-size long diffusion region, known as the Sweet-Parker solution \citep{sweet58a,parker57a}. The idea of a spatially localized anomalous resistivity was later invoked to explain the localization needed \citep{ugai77a,sato79a,kulsrud01a}, but no clear evidence of such anomalous resistivity during collisionless reconnection has yet been identified.

Kinetic simulations beyond the MHD model suggest that antiparallel reconnection with an open outflow geometry occurs when the current sheet thins down to the ion inertial scale \citep{jara-almonte21a,daughton09a,cassak05a,bhattacharjee04a}. When this occurs, the Hall term in the generalized Ohm's law \citep{vasyliunas75a,swisdak08a} dominates the electric field in the  ion diffusion region (IDR), where the ions become demagnetized. The correlation between the Hall effect and fast reconnection was clearly demonstrated in the GEM reconnection challenge study \citep{birn01a}; this study showed that simulation models with the Hall term in the generalized Ohm's law (particle-in-cell (PIC), hybrid and Hall-MHD) realize fast reconnection, while only the uniform resistive-MHD model, which lacks the Hall term, exhibits a slow rate \citep{sweet58a,parker57a}. However, it remains unclear why and how the Hall term localizes the diffusion region, producing an open geometry. The dispersive property of waves arising from the Hall term was proposed as an explanation \citep{mandt94a,shay99a,rogers01a,drake08a}, but the role of dispersive waves derived from linear analysis was called into question because reconnection can be fast even in systems that lack dispersive waves \citep{yhliu14a, tenbarge14a, stanier15a, bessho05a}. 

In this work, we illustrate the role of Hall physics in plasma energization and why this causes the open geometry necessary to achieve fast reconnection in electron-ion plasmas. The two key points are: (1) the Hall term ${\bf E}_{{\rm Hall}}={\bf J}\times{\bf B}/n{\rm e}c$, while it dominates the electric field within the IDR, does not convert energy into plasmas because ${\bf J}\cdot{\bf E}_{{\rm Hall}}={\bf J} \cdot ({\bf J}\times{\bf B})/n{\rm e}c=0$. Thus, the inflowing plasma only gains a small amount of thermal energy within the IDR. (2) Insufficient pressure buildup at the x-line, where the magnetic field lines change their connectivity, causes the upstream magnetic pressure to locally pinch the diffusion region, opening out the exhaust \citep{yhliu20a}. For reconnection of antiparallel magnetic fields, an open geometry occurs if $P|_{\rm xline} < B_{x0}^2/8{\rm \pi}+P_0$, where $P|_{\rm xline}$ is the thermal pressure at the x-line, $P_0$ is the asymptotic thermal pressure and $B_{x0}^2/8{\rm \pi}$ is the magnetic pressure based on the asymptotic magnetic field $B_{x0}$ far upstream from the IDR. These two results are used to derive a first-principles theory of the reconnection rate (the phrase ``{\it first-principles}'' refers to a theory that does not rely on measured empirical inputs from the simulations or observations).
In order to show this pressure depletion during magnetic reconnection in electron-ion plasmas, we use PIC simulations to investigate the role of Hall electromagnetic fields in energy conversion and kinetic heating near the x-line. The cross-scale coupling from the mesoscale upstream MHD region, the IDR, and down to the electron diffusion region (EDR) is treated to obtain a prediction of the reconnection rate. Finally, we extend the discussion to systems without the Hall term, including electron-positron (pair) plasmas and resistive-MHD reconnection, explaining why the former is fast while the latter does not have an open outflow and is slow. We show that the same theoretical approach leads to the Sweet-Parker scaling, and provides the reason of why Sweet-Parker reconnection has a system-size long diffusion region.

\section{Results}

%\subsection{The importance of heating near the x-line}
We use 2-D PIC simulations to illustrate the key features of energy conversion in the diffusion region. Details of the simulation setup are in the ``Methods'' section. The units used in the presentation include the ion cyclotron time $\Omega_{\rm ci}^{-1}$ $\equiv$ $({\rm e}B_{x0}/m_{\rm i}c)^{-1}$, the Alfv\'en speed $V_{\rm A0}$ $\equiv$ $B_{x0}/(4{\rm \pi} n_0 m_{\rm i})^{1/2}$ based on $B_{x0}$ and the background density $n_0$, and the ion and electron inertial length $d_s $ $\equiv$ $c/(4{\rm \pi} n_0 {\rm e}^2/m_s)^{1/2}$ for species $s$ $=$ $\rm i$ and $\rm e$, respectively. The ion to electron mass ratio is $m_{\rm i}/m_{\rm e}=400$ and the background plasma beta is $\beta=0.01$.

\begin{figure}
\includegraphics[width=8.5cm]{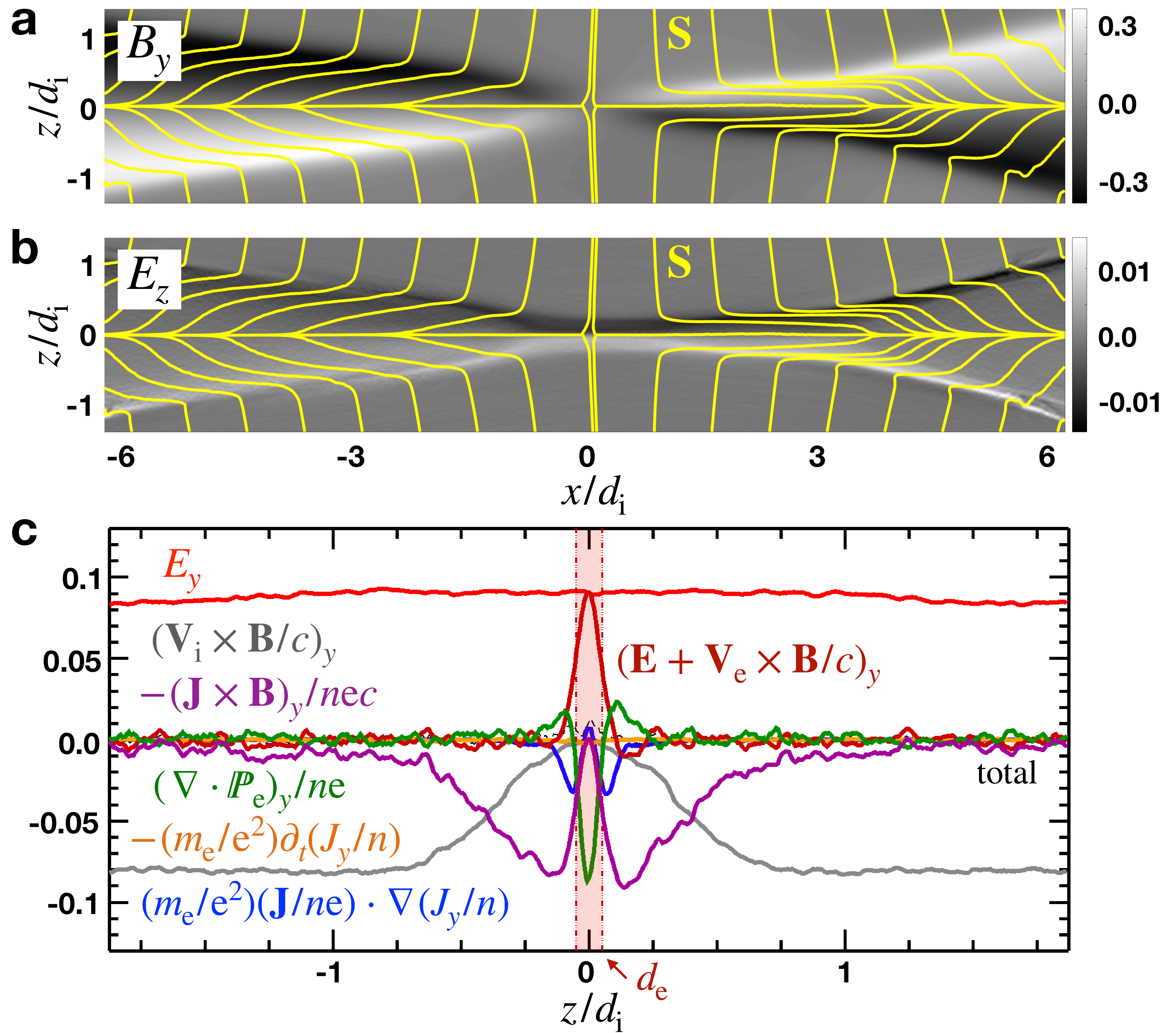} 
\caption {{\bf Hall electromagnetic fields and the generalized Ohm's law.} {\bf a}) The Hall magnetic field $B_y$ and {\bf b}) the Hall electric field $E_z$ (normalized by $B_{x0}$) overlaid with Poynting vector ${\bf S}$ streamlines (yellow) at time $48/\Omega_{\rm ci}$. {\bf c}) The out-of-plane component of terms in the generalized Ohm's law (normalized by $B_{x0}V_{\rm A0}/c$) across the x-line in the inflow direction. Variables ${\bf E}$, ${\bf B}$, ${\bf V}_{\rm i}$, ${\bf V}_{\rm e}$, ${\bf J}$, $\mathbb{P}_{\rm e}$, $n$, ${\rm e}$, $m_{\rm e}$ and $c$ are electric field, magnetic field, ion velocity, electron velocity, current density, electron pressure tensor, density, proton charge, electron mass and the speed of light, respectively. The vertical red transparent band marks the electron diffusion region (EDR). 
} 
\label{Hall}
\end{figure}

\subsection{The role of Hall electromagnetic fields}
Figure~\ref{Hall}a shows the out-of-plane magnetic field $B_y$ at time $48/\Omega_{\rm ci}$, which is the Hall quadrupole field within the IDR of magnetic reconnection in collisionless electron-ion plasmas {\cite{sonnerup79a}}. Importantly, this Hall quadrupole magnetic field $B_y$ along with the inward-pointing Hall electric field $E_z$, shown in Fig.~\ref{Hall}b, constitute a Poynting vector $S_x=-cE_z B_y/4{\rm \pi}$ in the $x$-direction. This component diverts the inflowing electromagnetic energy toward the outflow. This is shown by the streamlines of ${\bf S}=c{\bf E}\times{\bf B}/4{\rm \pi}$ in yellow, which bend in the $x$ direction significantly before reaching $z=0$.
These Hall electromagnetic fields arise from the Hall term in the generalized Ohm's law \cite{vasyliunas75a,swisdak08a,hesse11a}, ${\bf E}+{\bf V}_{\rm i} \times{\bf B}/c={\bf J}\times{\bf B}/n{\rm e}c-\nabla\cdot \mathbb{P}_{\rm e}/n{\rm e}+(m_{\rm e}/{\rm e}^2)d\left({\bf J}/n\right)/dt$ where $d/dt\equiv \partial_t - ({\bf J}/n{\rm e})\cdot\nabla$. The left-hand side (LHS) is the ideal electric field that becomes finite when the ion frozen-in condition is violated. Terms on the right-hand side (RHS) contribute to this violation in kinetic plasmas, including the Hall term, the electron pressure divergence term, and the electron inertia term. Fig.~\ref{Hall}c shows the terms in the out-of-plane ($y$) component of Ohm's law in a vertical cut through the x-line; the Hall term $({\bf J}\times {\bf B})_y/n{\rm e}c$ (in purple) is the dominant term supporting the reconnection electric field $E_y$ (in red) between the ion inertial scale $d_{\rm i}$ and the electron inertial scale $d_{\rm e}$. The Hall term arises because of the decoupling of the relatively immobile ions from the motion of electrons that remain frozen-in to the magnetic fields \citep{sonnerup79a}. Electrons, the primary current carrier within the IDR (i.e., ${\bf J} \simeq -{\rm e} n {\bf V}_{\rm e}$), then drag (both reconnected and not-yet reconnected) magnetic field lines out of the reconnection plane, producing the out-of-plane quadrupolar Hall magnetic field \citep{drake08a,mandt94a,ren05a,burch16a}.

\begin{figure}
\includegraphics[width=8cm]{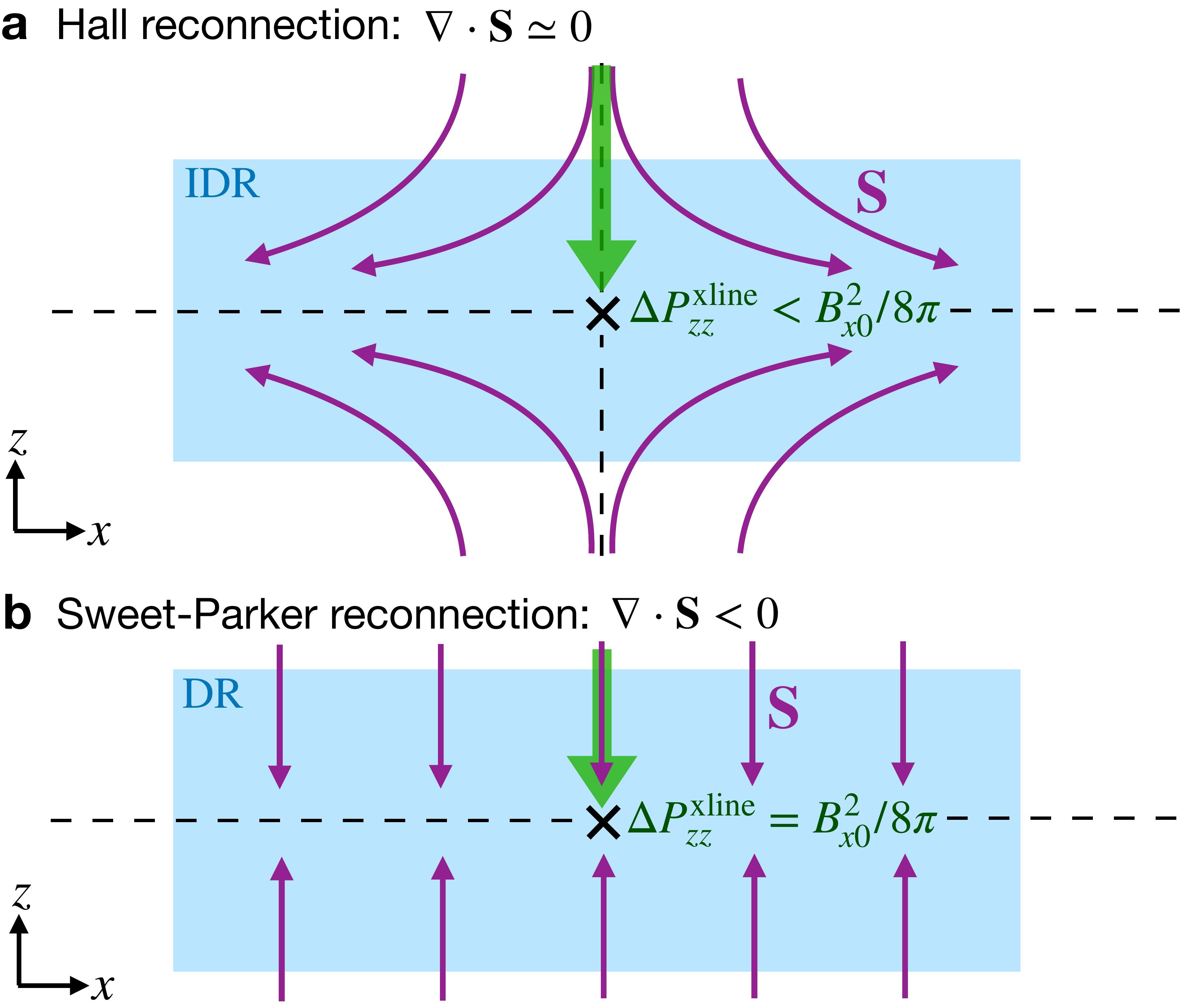} 
\caption {{\bf Transport patterns of electromagnetic energy in Hall reconnection and Sweet-Parker reconnection.} {\bf a}) Hall reconnection, where $\nabla\cdot {\bf S}\simeq -{\bf J}\cdot {\bf E}_{{\rm Hall}}= 0$ within the ion diffusion region (IDR). It requires a Poynting vector ${\bf S}$ streamline pattern that leads to an energy void around the x-line, limiting the difference in the $zz$-component of the pressure tensor between the x-line and the far upstream asymptotic region $\Delta P_{zz}^{\rm xline}$ (green arrow), and thus localizes the diffusion region. $B_{x0}^2/8{\rm \pi}$ is the asymptotic magnetic pressure far upstream from the IDR. {\bf b}) Sweet-Parker reconnection, where $\nabla\cdot {\bf S}< 0$ within the diffusion region (DR). The ${\bf S}$ streamlines thus can end uniformly on the outflow symmetry line, producing an elongated diffusion region.} 
\label{divS}
\end{figure}

Since the Hall term dominates the electric field ${\bf E}\simeq {\bf E}_{{\rm Hall}}$ $=$ $ {\bf J}\times {\bf B}/n{\rm e}c$ inside the IDR, then $\nabla\cdot {\bf S}$ $=$ $-{\bf J}\cdot {\bf E}$ $ \simeq 0$ per Poynting's theorem in the steady state. Along the inflow symmetry line ($x=0$) toward the x-line magnetic energy $B^2/8{\rm \pi}\rightarrow 0$ since $|B_x|$ decreases. Also, $B_z=0$ and $B_y=0$ (in antiparallel reconnection) due to symmetry. Consequentially, $\nabla\cdot {\bf S}$ $\simeq$ $0$ requires the ${\bf S}$ streamlines to be diverted to the outflow direction as illustrated in Fig.~\ref{divS}a (also shown in Fig.~\ref{Hall}a-b, consistent with the presence of $S_x$ $=$ $-cE_zB_y/4{\rm \pi}$). Since Poynting flux transports electromagnetic energy, this ${\bf S}$ streamline pattern implies an energy void centered around the x-line. This pattern introduces the localization to the diffusion region, even in an initially planar current sheet. In contrast, in resistive-MHD, $\nabla\cdot {\bf S}$ $=$ $-{\bf J}\cdot{\bf E}$ $\simeq$ $-\eta J_y^2$ $<$ $0$ where $\eta$ is the resistivity. Thus, as illustrated in Fig.~\ref{divS}b, ${\bf S}$ streamlines do not need to bend (i.e., $S_x\simeq 0$), instead ending and distributing energy uniformly on the outflow symmetry line ($z=0$).  This is why the diffusion region in Sweet-Parker reconnection is not localized.

To quantify the degree of localization, we need to estimate the thermal pressure at the x-line. The key is that ${\bf J}\cdot {\bf E}$ $\simeq$ $0$ inside the Hall dominated IDR limits the energy conversion to particles and thus also limits the difference in the $zz$-component of the pressure tensor between the x-line and the far upstream asymptotic region $\Delta P_{zz}^{\rm xline}\equiv P_{zz}|_{\rm xline}-P_0$ (illustrated in Fig.~\ref{divS}a). Given that magnetic pressure $B^2/8{\rm \pi}=0$ at the antiparallel reconnection x-line, if $\Delta P_{zz}^{\rm xline} < B_{x0}^2/8{\rm \pi}$, the reconnecting field bends toward the x-line as it approaches the x-line due to the force-balance condition $\nabla (P+B^2/8{\rm \pi})=({\bf B}\cdot \nabla){\bf B}/4{\rm \pi}$ \citep{yhliu20a}. This bending makes the outflow exhausts open out. This fact will be used to develop a first-principles theory of the reconnection rate.\\

\subsection{Non-vanishing kinetic heating within the IDR}
\begin{figure}
\includegraphics[width=9cm]{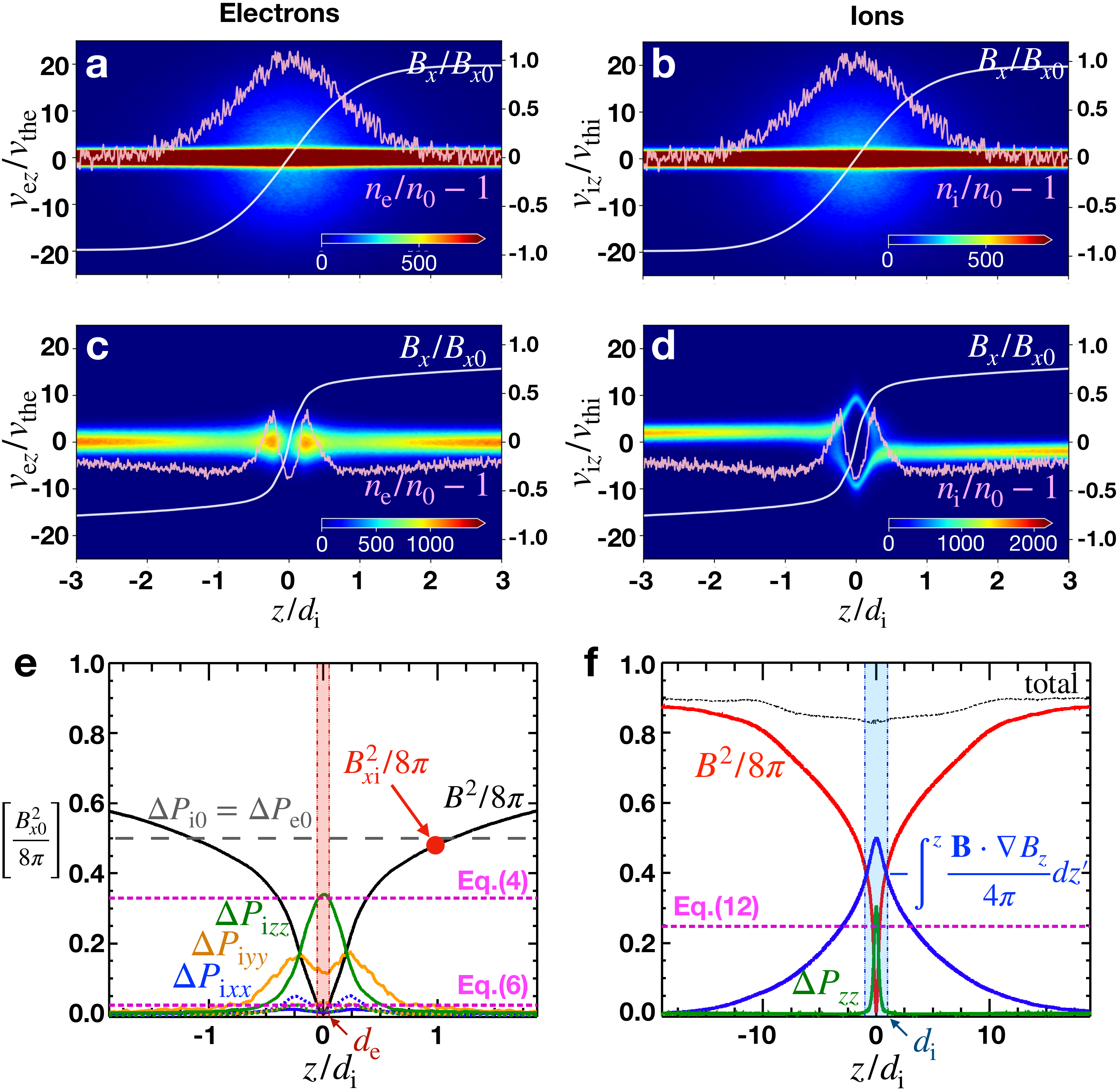} 
\caption {{\bf Phase space diagrams, thermal pressures, and force-balance.} The phase space diagrams (inflow velocity $v_z$ versus $z$-location) of the initial {\bf a}) electrons and {\bf b}) ions along the inflow symmetry line across the x-line at $(x,z)=(0,0)$. The structured but more tenuous {\bf c}) electrons and {\bf d}) ions in the same phase space at time $48/\Omega_{\rm ci}$. Particles are collected within $x/d_{\rm e}$ $\in$ $[-5,5]$ with velocities normalized to their background thermal speed $v_{{\rm th}s}$ for species $s={\rm e}$ and ${\rm i}$. Profiles of the magnetic field $B_x$ (white) and the density relative to the background value $n_s/n_0-1$ (pink) are overlaid for reference and their values are given by the right axis. The colormaps in panels {\bf a} and {\bf b} are capped to better show the hot current sheet components, thus the cold dense background plasmas appear as dark red at $v_{sz}\simeq 0$.
{\bf e}) The diagonal components of ion pressure relative to the far upstream value in solid lines ($\Delta P_{{\rm i}xx}, \Delta P_{{\rm i}yy}, \Delta P_{{\rm i}zz}$) at time $48/\Omega_{\rm ci}$ along the same inflow symmetry line, and that of electrons in dotted lines (near the bottom). For comparison, the initial scalar pressures at the x-line ($\Delta P_{\rm i0}$, $\Delta P_{\rm e0}$) are marked by the grey dashed horizontal line at value 0.5. The magnetic pressure at the ion-inertial ($d_{\rm i}$) scale, $B_{x{\rm i}}^2/8{\rm \pi}$, is indicted by the red dot.
{\bf f}) The $z$-component of the integrated forces in Eq.~(\ref{integrated_force}). The vertical red transparent band in panel {\bf e} marks the electron diffusion region (EDR), while the vertical blue transparent band in panel {\bf f} marks the ion diffusion region (IDR). Predictions from Eqs.~(\ref{Pizz}), (\ref{Bxe_Bxi}) and (\ref{Delta_Pzz}) are marked by the magenta horizontal lines in panels {\bf e} and {\bf f}.} 
\label{heating}
\end{figure}

Even though total plasma heating is limited within the IDR because ${\bf J}\cdot{\bf E}_{{\rm Hall}}=0$, there is a nonzero energy conversion arising from the convection electric field $-{\bf V}_{\rm i}\times {\bf B}/c$ (the grey curve in Fig.~\ref{Hall}c) that is critical for modeling the thermal pressure at the x-line. To reveal the kinetic heating process within the IDR along the inflow direction, we show phase space diagrams in Fig.~\ref{heating}. Figures~\ref{heating}a and \ref{heating}b show the initial (reduced) distributions $f(v_z,z)$ of electrons and ions, respectively. The density profiles are shown by $n_s/n_0-1$ in pink where $s={\rm e}$ and ${\rm i}$. The initial denser and hotter populations are visible within the current sheet ($|z|\lesssim1d_{\rm i}$); they balance the magnetic pressure across the initial Harris sheet. At time $48/\Omega_{\rm ci}$, the phase space diagrams of electrons and ions through the x-line show rich structures in Fig.~\ref{heating}c and \ref{heating}d, respectively. The pink profiles therein show that the density around the x-line ($z=0$) essentially matches the upstream value. On the other hand, while quasi-neutrality remains valid, ions in Fig.~\ref{heating}d form a phase space hole centered around the x-line. 

The ion distribution in Fig.~\ref{heating}d arises because the Hall electric field $E_z$ (Fig.~\ref{Hall}b) ballistically accelerates ions from both sides of the current sheet toward the x-line. They penetrate across the mid-plane, forming counter-streaming ion beams and thus this phase space hole \citep{wygant05a, LJChen08a, aunai11a}. Note that the acceleration is ``{\it ballistic}'' for inflowing ions because ions are already demagnetized within the IDR and they see the Hall $E_z$ as a DC field. Importantly, this kinetic effect increases the ion $zz$ pressure component above the asymptotic value, as quantified by $\Delta P_{{\rm i}zz}\equiv P_{{\rm i}zz}(z)-P_0$. Kinetically, the Hall $E_z$ arises from charge-separation due to the relative inflow motion between lighter, faster electrons and heavier, slower ions. This $E_z$ then speeds up ions and slows down electrons, which self-regulates its magnitude. Thus, the associated $\Delta P_{{\rm i}zz}$ buildup, even though effective, does not dominate the incoming energy budget during reconnection.

We note that the reconnection electric field $E_y$ is $\sim 6$ times smaller than the peak $E_z$ in our simulation, and it accelerates both species in the out-of-plane direction ($\pm y$ for ions and electrons, respectively). Some energy may be imparted into $\Delta P_{zz}$ through particle meandering motions, but the ballistic acceleration by $E_y$ is more efficient in imparting its energy into bulk kinetic energy of the current carriers in the $y$-direction \citep{hesse18a, hesse11a} rather than to thermal energy. Thus, it is not expected to greatly alter $\Delta P_{zz}$ and is ignored here.

In Fig.~\ref{heating}e, we show diagonal elements of the pressure tensor of electrons and ions (normalized to $B_{x0}^2/8{\rm \pi}$) through the x-line at time $48/\Omega_{\rm ci}$. 
For comparison, the initial pressure of each species, which completely balances the upstream magnetic pressure $B_{x0}^2/8{\rm \pi}$, is marked by the grey-dashed horizontal line (note that $\mathbb{P}_{\rm e}=\mathbb{P}_{\rm i}$ initially for this simulation). In the steady state, the ion $\Delta P_{{\rm i}zz}$ (solid green) is slightly reduced while the electron $\Delta P_{{\rm e}zz}$ (dotted green) is almost completely depleted. Other significant pressure depletion occurs in ion $\Delta P_{{\rm i}xx}$ (solid blue), electron $\Delta P_{{\rm e}xx}$ (dotted blue) and electron $\Delta P_{{\rm e}yy}$ (dotted orange). Ion $\Delta P_{{\rm i}yy}$ (solid orange) is roughly half of $\Delta P_{{\rm i}zz}$.

Our interest is in $P_{zz}$, since it affects force-balance in the inflow direction. Along the inflow, the ram pressure ($\sum_s^{{\rm i}, {\rm e}} n_sm_s V_{sz}^2$) is small, so the force-balance condition in the $z$ direction is $({\bf J}\times {\bf B})_z/c \simeq (\nabla \cdot \mathbb{P})_z$, where the total pressure $\mathbb{P}\equiv\sum_s^{{\rm i}, {\rm e}} \mathbb{P}_s$. Using ${\bf J}\times{\bf B}/c={\bf B}\cdot \nabla {\bf B}/4{\rm \pi}-\nabla B^2/8{\rm \pi}$ and integrating along the inflow ($z$) direction, the force-balance condition reads
\begin{equation}
\frac{B^2}{8{\rm \pi}}+\Delta P_{zz}-\int^z \frac{{\bf B}\cdot\nabla B_z}{4{\rm \pi}}dz'\simeq {\rm constant}.
\label{integrated_force}
\end{equation}
Initially, the magnetic pressure is totally balanced by the thermal pressure with no contribution from curvature. In the steady-state shown in Fig.~\ref{heating}f, the thermal pressure (green) at the x-line drops significantly, and the upstream magnetic tension (blue) at the mesoscale develops to counter balance the magnetic pressure (red). This tension force is realized through the bending of upstream field lines, which produces the Petschek-type open exhaust with a localized diffusion region \citep{yhliu20a,yhliu17a}. \\

%\subsection{First-principles calculation of the reconnection rate}

\subsection{Estimating the pressure within the IDR}
Within the IDR and outside the EDR, the electric field is ${\bf E}$ $=$ $-{\bf V}_{\rm e}\times{\bf B}/c$ $=$ $-{\bf V}_{\rm i}\times {\bf B}/c+{\bf J}\times {\bf B}/n{\rm e}c$. Thus the local energy conversion rate is 
\begin{equation}
{\bf J}\cdot {\bf E}=-{\bf J}\cdot {\bf V}_{\rm i}\times {\bf B}/c={\bf V}_{\rm i} \cdot {\bf J}\times {\bf B}/c={\rm e}n{\bf V}_{\rm i}\cdot {\bf E}_{{\rm Hall}}. 
\end{equation}
All of the work is done on the ions since the rate of work done on electrons is $-{\rm e}n{\bf V}_{\rm e}\cdot{\bf E}={\rm e}n{\bf V}_{\rm e}\cdot({\bf V}_{\rm e}\times{\bf B}/c)=0$. Note that even though ${\bf J}\cdot {\bf E}_{{\rm Hall}}=0$, the ions do gain energy by ${\bf E}_{{\rm Hall}}$, consistent with the ballistic acceleration displayed in Fig.~\ref{heating}d.

\begin{figure}
\includegraphics[width=8.5cm]{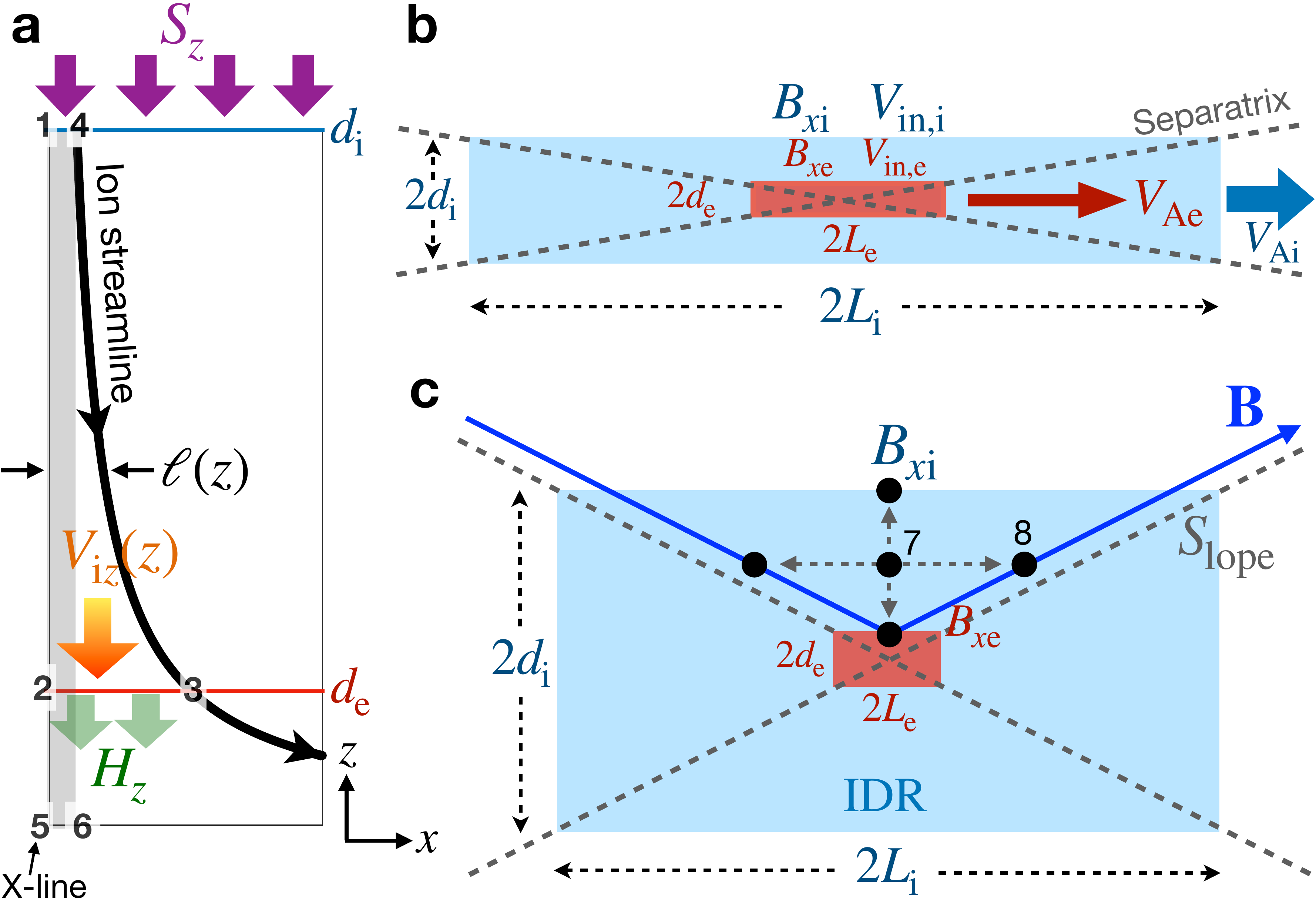} 
\caption {{\bf Diagrams of diffusion regions for theoretical modeling.} {\bf a}) The Gaussian surface (1-2-3-4) in the $\ell(z)\rightarrow 0$ limit that is used to calculate the ion pressure buildup $P_{{\rm i}zz}|^{d_{\rm e}}_{d_{\rm i}}$ between the electron inertial scale ($d_{\rm e}$) and ion inertial scale ($d_{\rm i}$). The x-line is located at the lower-left corner (point 5). $\ell(z)$ is the distance between the ion streamline and the inflow symmetry line (line 1-5) as a function of $z$.  The downward purple, green and orange arrows represent the incoming Poynting flux $S_z$, enthalpy flux $H_z$, and ion velocity $V_{{\rm i}z}$, respectively. {\bf b}) The two-scale diffusion region structure used to derive $B_{x{\rm e}}/B_{x{\rm i}}$. The blue box represents the ion diffusion region (IDR) and the red box represents the electron diffusion region (EDR). $V_{\rm in,i}$ ($V_{\rm in,e}$) is the ion (electron) inflow velocity at the ion (electron) inertial scale where the local magnetic field is $B_{x{\rm i}}$ ($B_{x{\rm e}}$). $V_{\rm Ai}$ and $V_{\rm Ae}$ are the ion Alfv\'en speed and electron Alfv\'en speed based on the local quantities, respectively. {\bf c}) The region used to derive the slope of the separatrix, $S_{{\rm lope}}$. The blue (red) box represents the IDR (EDR). The blue solid line depicts an upstream magnetic field $\bf B$ line adjacent to the separatrix shown by diagonal dashed lines.} 
\label{theories}
\end{figure}

Near the inflow symmetry line ($x=0$), $J_x$, $B_y$, $B_z$ and $V_{ix}$ are negligible because of the symmetry in antiparallel reconnection. Thus ${\bf J}\cdot {\bf E}\simeq -V_{{\rm i}z}J_yB_x/c+V_{{\rm i}y}J_zB_x/c$. The first term is work done by the Hall $E_z\simeq -J_yB_x/n{\rm e}c$, while the second term is work done by the reconnection electric field $E_y\simeq J_zB_x/n{\rm e}c$.   
Following the discussion of Fig.~\ref{heating}, the effective $\Delta P_{{\rm i}zz}$ buildup primarily arises from ballistic acceleration by the Hall $E_z$ that results in counter-streaming ions near the x-line. We calculate the contribution of $-V_{{\rm i}z}J_yB_x/c$ to the total rate of energy conversion $\int {\bf J}\cdot {\bf E} dV$ for the volume enclosed by the Gaussian surface (1-2-3-4) in Fig.~\ref{theories}a. The right surface (3-4) is chosen to coincide with an ion streamline near the inflow symmetry line, so there is no ion energy flux through that surface. The $x$-location of this surface relative to $x=0$ is quantified by $\ell(z)$. Then  
\begin{equation}
\begin{split}
\int_{1234} -\frac{V_{{\rm i}z}J_yB_x}{c} dV &=\int_{1234} -V_{{\rm i}z}\frac{c}{4{\rm \pi}}\left(\frac{\partial B_x}{\partial z}\right)\frac{B_x}{c} dxdydz\\
 & \simeq \ell_y\int_{d_{\rm e}}^{d_{\rm i}}\ell(z)V_{{\rm i}z}(z)\frac{\partial}{\partial z}\left(\frac{B_x^2}{8{\rm \pi}}\right) dz \\
 & \simeq \left(\frac{B_{x{\rm i}}^2-B_{x{\rm e}}^2}{8{\rm \pi}}\right)V_{{\rm i}z}(d_{\rm i})\ell(d_{\rm i})\ell_y. \\
\end{split}
\label{JdotE}
\end{equation}
Here we assumed $|\partial_x B_z|$ $\ll$ $|\partial_zB_x|$ in writing $J_y$ $\simeq$ $(c/4{\rm \pi})\partial_z B_x$. 
We take the $\ell(z) \rightarrow 0$ limit, so the variables within the integrand are approaximately independent of $x$. We defined $B_{x{\rm i}}\equiv B_{x}|_{d_{\rm i}}$ and $B_{x{\rm e}}\equiv B_x|_{d_{\rm e}}$, and used $\ell_y$ as the dimension of the integrated surface in the out-of-plane (translationally invariant) direction. We also assumed that the density is nearly incompressible along the inflow so $V_{{\rm i}z}(z)\ell(z)\simeq V_{{\rm i}z}(d_{\rm i})\ell(d_{\rm i})$.

From the discussion of Fig.~\ref{heating}d, we expect that most of this energy is converted to $P_{{\rm i}zz}$, which contributes to the ion enthalpy flux ${\bf H}_{\rm i}$ $=$ $(1/2)\mbox{Tr}(\mathbb{P}_{\rm i}) {\bf V}_{\rm i}$ $+$ $\mathbb{P}_{\rm i}\cdot {\bf V}_{\rm i}$ \cite{birn10b} that enters the EDR in the vicinity of the x-line. The associated net enthalpy flux difference $\oint{\bf H}\cdot d{\bf A}$ through the Gaussian surface 1-2-3-4 is $(3/2)P_{{\rm i}zz}|_{d_{\rm e}}V_{{\rm i}z}(d_{\rm e})\ell(d_{\rm e})\ell_y$ $-$ $(3/2)P_{{\rm i}zz}|_{d_{\rm i}}V_{{\rm i}z}(d_{\rm i})\ell(d_{\rm i})\ell_y$ $\simeq$ $(3/2)P_{{\rm i}zz}|_{d_{\rm i}}^{d_{\rm e}} V_{{\rm i}z}(d_{\rm i})\ell(d_{\rm i})\ell_y$, where $P_{{\rm i}zz}|_{d_{\rm i}}^{d_{\rm e}}$ $\equiv$ $P_{{\rm i}zz}|_{d_{\rm e}}$ $-$ $P_{{\rm i}zz}|_{d_{\rm i}}$. Equating this quantity with the RHS of Eq.~(\ref{JdotE}) gives
\begin{equation}
P_{{\rm i}zz}|_{d_{\rm i}}^{d_{\rm e}} \simeq \frac{2}{3}\left(\frac{B_{x{\rm i}}^2-B_{x{\rm e}}^2}{8{\rm \pi}}\right). 
\label{Pizz}
\end{equation}
Since no work is done on electrons outside the EDR, $P_{{\rm e}zz}|_{d_{\rm i}}^{d_{\rm e}}$ $\simeq$ $0$ and thus the total thermal pressure difference $P_{zz}|_{d_{\rm i}}^{d_{\rm e}}$ $\simeq$ $P_{{\rm i}zz}|_{d_{\rm i}}^{d_{\rm e}}$. 
This thermal pressure increase is smaller than the magnetic pressure drop $(B_{x{\rm i}}^2-B_{x{\rm e}}^2)/8{\rm \pi}$ between the $d_{\rm i}$- and $d_{\rm e}$-scale, so there is insufficient pressure to balance forces in the $z$-direction without the bending of field lines, and Hall reconnection opens into a Petschek-type geometry. This predicted value of Eq.~(\ref{Pizz}), calculated using the measured $B_{x{\rm i}}$ and $B_{x{\rm e}}$, is plotted as a horizontal magenta line in Fig.~\ref{heating}e and compares well with the measured $P_{{\rm i}zz}|_{d_{\rm i}}^{d_{\rm e}}$ (green). \\

\subsection{Available magnetic energy at the EDR scale}
In order to estimate the relative magnetic pressure (energy) at the EDR, we write 
\begin{equation}
\frac{c E_{y{\rm i}}}{B_{x{\rm i}} V_{\rm Ai}}= \frac{V_{\rm in,i}}{V_{\rm Ai}}\simeq \frac{d_{\rm i}}{L_{\rm i}}\sim \frac{d_{\rm e}}{L_{\rm e}}\simeq \frac{V_{\rm in,e}}{V_{\rm Ae}}= \frac{c E_{y{\rm e}}}{B_{x{\rm e}}V_{\rm Ae}}.
\label{two_scale_derivation}
\end{equation}
The quantities are defined and illustrated in Fig.~\ref{theories}b. The first and last equalities come from the frozen-in conditions $E_{ys}=V_{{\rm in},s}B_{xs}/c$ at the inflow edges of the IDR and EDR for $s={\rm i}$ and ${\rm e}$, respectively. We use incompressibility for the second and fourth equalities. For the third equality, we use a geometrical argument that the magnetic field line threading the x-line and the corners of the EDR and the IDR is approximately straight, resulting in a similar aspect ratio for the EDR and the IDR.
At the ion-scale, the outflow speed is the ion Alfv\'en speed $V_{\rm Ai}\equiv B_{x{\rm i}}/(4{\rm \pi} n m_{\rm i})^{1/2}$. In contrast,  the electron outflow speed is the electron Alfv\'en speed based on the local conditions, $V_{\rm Ae}\equiv B_{x{\rm e}}/(4{\rm \pi} n m_{\rm e})^{1/2}$, since ions decouple from the motion of magnetic field lines in the electron-scale inside the IDR \citep{shay98a}.

By equating the first and last terms and noting that $E_y$ is uniform in 2D steady-state per Faraday's law (seen in Fig.~\ref{Hall}c), we find  
\begin{equation}
\frac{B_{x{\rm e}}^2}{B_{x{\rm i}}^2} \simeq \left(\frac{m_{\rm e}}{m_{\rm i}}\right)^{1/2}.
\label{Bxe_Bxi}
\end{equation}
Note that the equality between the first and the last terms is consistent with the high-cadence observation of Magnetospheric Multiscale Mission (MMS)  \citep{burch20a}. For $m_{\rm i}/m_{\rm e}=400$ as in the simulation, $B_{x{\rm e}}^2/B_{x{\rm i}}^2\simeq 0.05$. The predicted $B_{x{\rm e}}^2/8{\rm \pi}$ based on the measured $B_{x{\rm i}}^2/8{\rm \pi}$ in Fig.~\ref{heating}e compares well with the small $B_{x}^2/8{\rm \pi}$ ($\simeq B^2/8{\rm \pi}$ in black) value at the $d_{\rm e}$-scale. For the real proton to electron mass ratio $m_{\rm i}/m_{\rm e}=1836$, $B_{x{\rm e}}^2/B_{x{\rm i}}^2\simeq 0.023$. The smallness of $B_{x{\rm e}}^2/B_{x{\rm i}}^2$ makes the contribution of the pressure depletion within the EDR negligible. However, this imbalanced pressure becomes critical in pair plasmas where the EDR is the same as the IDR, as discussed later.\\

\subsection{Cross-scale coupling and the rate prediction}
To predict the reconnection rate, we use the force-balance condition $\nabla B^2/8{\rm \pi} +\nabla \cdot \mathbb{P}={\bf B}\cdot \nabla {\bf B}/4{\rm \pi}$ and geometry to couple the solutions at the IDR, EDR and the upstream MHD region. 
First, we discretize this equation at point {\bf 7} of Fig.~\ref{theories}c. In the $z$-direction,
\begin{equation}
\frac{B_{x{\rm i}}^2-B_{x{\rm e}}^2}{8{\rm \pi} (d_{\rm i}-d_{\rm e})}-\frac{P_{zz}|^{d_{\rm e}}_{d_{\rm i}}}{d_{\rm i}-d_{\rm e}}\simeq \left(\frac{B_{x{\rm i}}+B_{x{\rm e}}}{2}\right)\frac{2 B_{z8}}{4{\rm \pi} L_{\rm i} (d_{\rm i}-d_{\rm e})/d_{\rm i}}.
\label{force_balance}
\end{equation}
From geometry, the slope of the separatrix $S_{{\rm lope}}$ $ \simeq$ $ d_{\rm i}/L_{\rm i} $ $\simeq$ $B_{z8}/[(B_{x{\rm i}}+B_{x{\rm e}})/2]$. Solving for $S_{{\rm lope}}$ gives
\begin{equation}
S_{{\rm lope}}^2 \simeq \frac{B_{x{\rm i}}^2-B_{x{\rm e}}^2}{(B_{x{\rm i}}+B_{x{\rm e}})^2}-\frac{8{\rm \pi} P_{zz}|^{d_{\rm e}}_{d_{\rm i}}}{(B_{x{\rm i}}+B_{x{\rm e}})^2}.
\label{S_general}
\end{equation}
Physically, this expression relates the opening angle of the separatrix to the thermal pressure difference. Plugging in $P_{zz}|_{d_{\rm i}}^{d_{\rm e}}$ using Eq.~(\ref{Pizz}), we get
\begin{equation}
S_{{\rm lope}}^2\simeq \frac{1}{3}\left[\frac{1-(B_{x{\rm e}}/B_{x{\rm i}})}{1+(B_{x{\rm e}}/B_{x{\rm i}})}\right].
\label{S}
\end{equation} 

A similar analysis on the force-balance across the EDR gives 
\begin{equation}
S_{{\rm lope}}^2 \simeq 1-\frac{8{\rm \pi} P_{zz}|^{0}_{d_{\rm e}}}{B_{x{\rm e}}^2},
\label{S_EDR}
\end{equation}
where the slope of the separatrix is assumed similar inside and outside the EDR because the magnetic tension straightens out the field lines.

We now relate $B_{x{\rm i}}$ back to the upstream asymptotic field $B_{x0}$ using 
\begin{equation}
\frac{B_{x{\rm i}}}{B_{x0}}\simeq \frac{1-S_{{\rm lope}}^2}{1+S_{{\rm lope}}^2},
\label{Bxi_Bx0}
\end{equation}
which was previously derived using the force-balance condition upstream of the IDR at the mesoscale \cite{yhliu17a}. We again assume the slope of the separatrix is similar inside and outside the IDR. 

Finally, using Eqs.~(\ref{Pizz}), (\ref{S_EDR}) and (\ref{Bxi_Bx0}), we can rewrite the total thermal pressure buildup $\Delta P_{zz}^{\rm xline} \simeq P_{zz}|_{d_{\rm i}}^{d_{\rm e}}+P_{zz}|_{d_{\rm e}}^0$ in terms of the upstream asymptotic magnetic pressure $B_{x0}^2/8{\rm \pi}$ as
\begin{equation}
\Delta P_{zz}^{\rm xline} \simeq \left[\frac{2}{3}+\left(\frac{1}{3}-S_{{\rm lope}}^2\right)\frac{B_{x{\rm e}}^2}{B_{x{\rm i}}^2}\right]\left(\frac{1-S_{{\rm lope}}^2}{1+S_{{\rm lope}}^2}\right)^2\frac{B_{x0}^2}{8{\rm \pi}}.
\label{Delta_Pzz}
\end{equation}  
Using Eqs.~(\ref{Delta_Pzz}), (\ref{S}) and (\ref{Bxe_Bxi}), for $m_{\rm i}/m_{\rm e}=400$ we get $\Delta P_{zz}^{\rm xline}$ $\simeq$ $0.283 (B_{x0}^2/8{\rm \pi})$. This prediction is within $\simeq 20\%$ of the simulated $\Delta P_{zz}^{\rm xline}/(B_{x0}^2/8{\rm \pi})$ as indicated by the magenta horizontal line in Fig.~\ref{heating}f. The predicted $\Delta P_{zz}^{\rm xline}$ is considerably less than $B_{x0}^2/8{\rm \pi}$, which from Eq.~(\ref{integrated_force}) is consistent with there being an open outflow geometry.

Once the separatrix slope is determined, we can also obtain a first-principles prediction of the reconnection rate using the $R-S_{{\rm lope}}$ relation in Liu et al. \cite{yhliu17a}, 
\begin{equation}
R=S_{{\rm lope}}\left(\frac{1-S_{{\rm lope}}^2}{1+S_{{\rm lope}}^2}\right)^2\sqrt{1-S_{{\rm lope}}^2}, 
\label{R}
\end{equation}
which is $0.157$ for $m_{\rm i}/m_{\rm e}=1836$ and $0.172$ for $m_{\rm i}/m_{\rm e}=400$, consistent with the measured rate on the order of $\mathcal{O}(0.1)$ as shown in Fig.~\ref{Hall}c and literature \citep{birn01a,shay99a,hesse99a}.\\ 

\subsection{Pair Plasmas and Resistive-MHD}
In systems without the Hall effect, such as pair plasmas and resistive-MHD, energy conversion inside the diffusion region is totally different. There is no two-scale structure to the diffusion region. Thus, for the following discussion, we denote the microscopic thickness of the diffusion region as $d_{\rm m}$ (corresponding to $d_{\rm i}$ in Fig.~\ref{theories}a) and define $\ell_{\rm in}\equiv\ell(d_{\rm m})$, $V_{\rm in}\equiv |V_{{\rm i}z}(d_{\rm m})|$, $B_{x{\rm m}}\equiv B_{x}(d_{\rm m})$ and the inflowing Poynting vector $S_{\rm in}\equiv |S_z(d_{\rm m})|$. We consider the energy conversion rate integrated over the gray rectangular box (1-5-6-4) in Fig.~\ref{theories}a in the $\ell_{\rm in}\rightarrow 0$ limit to illustrate the critical difference compared to Hall reconnection.
Near the inflow symmetry line ($x=0$) in antiparallel reconnection, $J_x=0$, thus ${\bf J}\cdot {\bf E}$ $=$ $J_y E_y+J_z E_z$. Without the Hall effect, $E_z$ vanishes while $E_y$ is uniform in 2D steady-state, thus
\begin{equation}
\begin{split}
\int_{1564} {\bf J}\cdot {\bf E} dV & \simeq \ell_y\ell_{\rm in}E_y\int_0^{d_{\rm m}} \frac{c}{4{\rm \pi}}\left(\frac{\partial B_x}{\partial z}\right)dz \\
& =c\frac{E_yB_{x{\rm m}}}{4{\rm \pi}}\ell_{\rm in}\ell_y
=S_{\rm in}\ell_{\rm in}\ell_y.
\end{split}
\label{rectangular}
\end{equation}
Here we assumed $|\partial_x B_z| \ll |\partial_zB_x|$ (i.e., weak localization) again, so $J_y\simeq (c/4{\rm \pi})\partial_z B_x$.
Importantly, the last equality of Eq.~(\ref{rectangular}) indicates that the Poynting flux entering the top surface (1-4) is all converted to plasma energy within this narrow rectangular box. In other words, no Poynting flux is diverted to the outflow direction, as illustrated in Fig.~\ref{divS}b. The outflowing enthalpy flux $H_x$ and the bulk flow kinetic energy flux ${\bf K}\equiv\sum_s^{{\rm i}, {\rm e}}(1/2)nm_s V_s^2 {\bf V}_s$ \citep{birn10b,eastwood20a} in the $x$-direction $K_x$ compete for the inflowing energy. For example, in the low background-$\beta$ limit where both $H_z$ and $K_z$ at the inflow surface (1-4) are negligible, the energy conversion is $S_{\rm in}\ell_{\rm in}$ $\simeq$ $\int_{6}^4 H_x dz$ $+$ $\int_{6}^4K_x dz$. 

In electron-positron ($m_{\rm i}=m_{\rm e}$) pair plasmas \cite{bessho05a,yhliu17a,yhliu15a}, near the inflow symmetry line $K_x$ primarily comes from the bulk flow kinetic energy of the current carriers $\sum_s^{{\rm i}, {\rm e}}(1/2)n m_s V_{sy}^2 V_{sx}$. In the magnetically-dominated relativistic regime \cite{FGuo20a}, to sustain the extreme current density, $K_x$ can be large and it limits the energy available for the enthalpy $H_x$. This leads to a depleted pressure at the x-line and fast reconnection \cite{yhliu20a}. A similar competition could occur in non-relativistic low-$\beta$ pair plasmas. Without the inward-pointing Hall $E_z$, the counter-streaming ions that efficiently build up $\Delta P_{zz}$ in Hall reconnection are absent. The only other potential source of heating is through the reconnection electric field $E_y$, which efficiently increases the current carrier drift speed, but not $\Delta P_{zz}$ because the acceleration is primarily in the $y$-direction. Therefore, $\Delta P_{zz}$ should be less than $B_{x0}^2/8{\rm \pi}$ and fast reconnection with an open outflow should occur.

For (isotropic) resistive-MHD, $K_x$ near the inflow symmetry line vanishes because the current in MHD is not associated with any kinetic energy. All the inflowing energy is then converted into enthalpy. Using $E_y=V_{\rm in}B_{x{\rm m}}/c$ in Eq.~(\ref{rectangular}) then $S_{\rm in}\ell_{\rm in}$ $=$ $(B_{x{\rm m}}^2/4{\rm \pi}) V_{\rm in} \ell_{\rm in}$ $\simeq$ $\int_{6}^4 H_x dz $ $=$ $(5/2) \int_{6}^4 PV_{x} dz $ $<$ $(5/2)P|_6 \int_{6}^4V_{x} dz$, where the inequality arises from a reasonable thermal pressure profile that peaks at point {\bf 6} on the 4-6 line. In the incompressible limit, $\int_{6}^4V_{x} dz$ $=$ $V_{\rm in} \ell_{\rm in}$, so we get $P|_6$ $>$ $(4/5) B_{x{\rm m}}^2/8{\rm \pi}$, indicating that a balanced-pressure $P|_4^6$ $=$ $B_{x{\rm m}}^2/8{\rm \pi}$ becomes possible. Numerical simulations confirms that the thermal pressure at the x-line nearly balances the upstream magnetic pressure in Sweet-Parker reconnection. This implies $S_{{\rm lope}}\rightarrow 0$ from Eq.~(\ref{S_general}) and explains why Sweet-Parker reconnection is slow; note that the Sweet-Parker theory \cite{sweet58a,parker57a} itself does not address why the diffusion region length extends to the system size (i.e., $S_{{\rm lope}}\rightarrow 0$), but follows naturally in the present model.

In fact, we can recover the Sweet-Parker scaling using the framework laid out here. 
Using $E_y$ $=$ $\eta J_{y}|_{\rm xline}$ $\simeq$ $\eta (c/4{\rm \pi})(B_{x{\rm m}}/d_{\rm m})$ in Eq.~(\ref{rectangular}), we get
\begin{equation}
\begin{split}
\int_{1564} {\bf J}\cdot {\bf E} dV\simeq \eta \left(\frac{c}{4{\rm \pi}}\right)^2\frac{B_{x{\rm m}}^2}{d_{\rm m}}\ell_{\rm in}\ell_y.
\end{split}
\end{equation}
Since this energy is all converted into the enthalpy $(5/2)\ell_y\int_6^4 P V_x dz$ $\simeq$ $(5/2) \langle P \rangle \ell_y\int_6^4 V_x dz$ $\simeq$ $ (5/2) \langle P \rangle V_{\rm in}\ell_{\rm in}\ell_y$, we can solve for the average pressure $\langle P \rangle$. The pressure buildup can be estimated as 
\begin{equation}
\begin{split}
P|_{d_{\rm m}}^{0}\simeq \langle P \rangle \simeq \frac{4}{5}\left(\frac{L_{\rm i}}{d_{\rm m}}\right)^2\left(\frac{\eta c^2}{4{\rm \pi} V_{\rm Ai} L_{\rm i}}\right)\frac{B_{x{\rm m}}^2}{8{\rm \pi}},
\end{split}
\end{equation}
where we used plasma continuity $V_{\rm in}$ $\simeq$ $V_{\rm Ai} d_{\rm m}/L_{\rm i}$ over the entire diffusion region. Using this $P|_{d_{\rm m}}^0$ in Eq.~(\ref{S_EDR}) though noting $P_{zz}$ $=$ $P$ since the pressure is isotropic, $d_{\rm e}=d_{\rm i}$ $=$ $d_{\rm m}$, $B_{x{\rm e}}=B_{x{\rm i}}=B_{x{\rm m}}$ and using $L_{\rm i}/d_{\rm m}$ $\simeq$ $S_{{\rm lope}}$ as before, we get 
\begin{equation}
\begin{split}
S_{{\rm lope}}^2\simeq 1- \frac{4}{5}\left(\frac{1}{S_{{\rm lope}}}\right)^2\left(\frac{\eta c^2}{4{\rm \pi} V_{\rm Ai} L_{\rm i}}\right).
\end{split}
\end{equation}
Looking for a solution in the $S_{{\rm lope}}\rightarrow 0$ limit, as justified above, we get the reconnection rate from Eq.~(\ref{R}) to be
\begin{equation}
\begin{split}
R_{\rm SP} \simeq S_{{\rm lope}} \simeq \sqrt{\frac{4}{5}\left(\frac{\eta c^2}{4{\rm \pi} V_{\rm Ai} L_{\rm i}}\right)}.
\end{split}
\label{R_SP}
\end{equation}
Up to a numerical factor near unity, this result is the Sweet-Parker scaling \citep{sweet58a,parker57a} of the reconnection rate in uniform resistivity MHD. 

To see the compatibility of the model developed here with the result of a spatially localized resistivity in MHD \citep{ugai77a,sato79a,biskamp86a,uzdensky00a,SCLin21a}, one needs to retain $-\partial_x B_z$ in the estimate of $J_y$. This leads to $\int_{1564} {\bf J}\cdot {\bf E} dV< S_{\rm in}\ell_{\rm in}\ell_y$, instead of them being equal as in Eq.~(\ref{rectangular}). Physically, a finite $\partial_x B_z$ indicates a localized diffusion region and gives rise to an $x$ component of the Poynting vector $cE_y B_z/4{\rm \pi}$.
This $S_x$ also diverts the inflowing energy to the outflow direction, analogous to $-cE_zB_y/4{\rm \pi}$ arising from the Hall effect (Fig.~\ref{Hall}). Hence, a localized diffusion region with a depleted pressure at the x-line takes place. However, in this case, the diffusion region localization is introduced by hand or some unidentified mechanisms \cite{kulsrud01a}. 

\section{Conclusions}
We have shown that, counter-intuitively, a lower energy conversion rate ${\bf J}\cdot{\bf E}$ along the inflow toward the x-line makes reconnection faster because the lower pressure requires upstream magnetic field lines to bend to enforce force balance, therefore opening the outflow exhaust.
We predict that the high thermal pressure required to get an elongated planar current sheet is not energetically sustainable at the x-line of electron-proton plasmas because ${\bf J}\cdot {\bf E}_{{\rm Hall}}$$=0$.  
A significant portion of incoming electromagnetic energy is not transported to the x-line, but is diverted to the outflow by Hall fields. 
The theory presented here directly links the Hall effect to diffusion region localization, and the linkage is the pressure depletion at the x-line. 
The predicted pressure drop compares well with simulations of collisionless magnetic reconnection. Through cross-scale coupling between the EDR, IDR and the upstream MHD region, the fast reconnection rate of order $\mathcal{O}(0.1)$ is derived (Eqs.~(\ref{R}),(\ref{S}),(\ref{Bxe_Bxi})) from first-principles for the first time to the best of our knowledge. A closer agreement can be made after considering the reconnection outflow speed reduction by thermal pressure effects \citep{XLi21a,haggerty18a,yhliu12a}; a predicted $R\simeq 0.075$ can be read off from the $R-S_{{\rm lope}}$ relation (similar to Eq.~(\ref{R})) in Fig. 5(c) of Li and Liu \cite{XLi21a} using the same $S_{{\rm lope}}$, which is in excellent quantitative agreement with the simulated $E_y$ here in Fig.~\ref{Hall}c. In addition, the competition between different forms of energy flux explains why Sweet-Parker reconnection does not have an open exhaust and is slow, while reconnection in electron-positron (pair) plasma is fast. The same theoretical framework recovers the Sweet-Parker scaling law (Eq.~(\ref{R_SP})). 

This work is dedicated to explaining the primary localization mechanism for a stable single x-line in collisionless plasmas relevant to magnetospheric, solar and laboratory applications. If a stable single x-line can be realized (i.e., the steadiness is justified by the nearly uniform $E_y$ and negligible $\partial_t (J_y/n)$ shown in Fig.~\ref{Hall}c), the open outflow geometry can suppress the generation of secondary tearing modes \citep{CShi19a, shepherd10a, daughton09a} and time-dependent dynamics become less important. 
However, if the localization of a single x-line is weak, the reconnecting layer is very thin, or pressure-balance across the opened outflow exhausts can not be established \citep{yhliu20a}, then secondary tearing modes will be triggered. A cycle of fast generation and ejection of secondary tearing islands provides additional localization mechanism to increase the average reconnection rate \citep{yhliu20a,comisso16b,pucci14a,YMHuang10a,loureiro07a,daughton07a}. 

Finally, the Hall effect arises whenever the current sheet thins down to the ion kinetic scale, thus even in a thin 3D current sheet, the argument based on the two key points in the ``Introduction'' still holds. Large-scale 3D PIC simulations also suggest that broader turbulent current sheets can collapse to thin reconnecting layers in the kinetic scale \citep{FGuo21a}. Kinetic reconnecting layers also persist and dominate a current sheet that is filled with self-generated turbulence \citep{daughton11a, daughton14a, yhliu14a, le18a, stanier19a, price16a}. Importantly, the reconnection process is often quasi-2D in nature. This has been ascertained from the abundance of data from the MMS mission, which has unprecedented spatial and temporal resolutions. In particular, 2D simulations have done an excellent job reproducing detailed reconnection dynamics \citep{torbert18a, RNakamura18a, TKMNakamura18a, genestreti18b, burch20a}. 
Nevertheless, it remains an open question to explore whether fast reconnection can proceed in Nature without eventually forming a dominating kinetic current sheet in three-dimensional plasmas.

\section{Methods}
We carry out 2D PIC simulations of magnetic reconnection in a low background $\beta$ $\equiv$ $8{\rm \pi} P_0/B_{x0}^2$ plasma. The simulations are performed using the VPIC code~\citep{bowers08a}, which solves Maxwell's equations and the relativistic Vlasov equation. The simulation employs the Harris current sheet equilibrium \citep{harris62} that has the initial magnetic profile ${\bf B}$ $=$ $B_{x0}\tanh(z/\lambda)\hat{x}$ and the density profile $n$ $=$ $n_0\mbox{sech}^2(z/\lambda)$ $+$ $n_0$; here $\lambda$ $=$ $1d_{\rm i}$ is the initial half-thickness of the current sheet, where $d_{\rm i}$ $\equiv$ $c/(4{\rm \pi} n_0 {\rm e}^2/m_{\rm i})^{1/2}$ is the ion-inertial scale based on the background plasma density $n_0$, which is also the peak density of the current sheet population in our setup; this choice avoids current sheet expansion due to density depletion as has been seen \citep{bessho10b,PWu11a}. 
The simulation size is $L_x\times L_z = 76.8d_{\rm i}\times38.4d_{\rm i}$ that spans the domain $[-L_x/2,L_x/2] \times [-L_z/2,L_z/2]$ with $n_x\times n_z = 12288\times 6144$ cells. There are $\simeq 15$ billion macro-particles. The $x$-direction boundaries are periodic, while the $z$-direction boundaries are conducting for fields and reflecting for particles. We use the ion-to-electron mass ratio $m_{\rm i}/m_{\rm e}=400$, the temperature ratio $T_{\rm i}/T_{\rm e}=1$, the plasma $\beta= 0.01$ based on the background temperature $T_0$ and background density $n_0$, and $\omega_{\rm pe}/\Omega_{\rm ce}=4$, where the electron plasma frequency $\omega_{\rm pe}=(4{\rm \pi} n_0 {\rm e}^2/m_{\rm e})^{1/2}$ and the electron cyclotron frequency $\Omega_{\rm ce}={\rm e} B_{x0}/m_{\rm e} c$. These result in the background electron thermal speed $v_{\rm the} \equiv (T_0/m_{\rm e})^{1/2}= 0.0125c$, ion thermal speed $v_{\rm thi} \equiv (T_0/m_{\rm i})^{1/2}= 0.000625c$ and Alfv\'en speed $V_{\rm A0}\equiv B_{x0}/(4{\rm \pi} n_0m_{\rm i})^{1/2}=0.0125c$, well within the non-relativistic regime for solar and magnetospheric applications. A localized initial magnetic field perturbation of amplitude $\delta B_z=0.03B_{x0}$ is added to induce single x-line reconnection at the center of the simulation domain.
To reduce noise in the generalized Ohm's law analysis, the presented data is time-averaged over an interval $0.085/\Omega_{\rm ci}$, where $\Omega_{\rm ci}={\rm e} B_{x0}/m_{\rm i} c$ is the ion cyclotron frequency.

\section{Data availability}
Access to the simulation data and scripts used to plot the figures are available at Zenodo (https://doi.org/10.5281/zenodo.6258255). All other data are available from the corresponding author on reasonable request.\\

\section{Code availability}
The simulations are conducted using an open source Vector Particle-In-Cell (VPIC) project, which is available at https://github.com/lanl/vpic. The simulation data are analyzed using IDL, Matlab and Python. The scripts are available at the data storage site.

\section{References}
%\bibliography{paper}


\section{Acknowledgments}
Y.L. is grateful for supports from NSF grant PHY- 1902867 through the NSF/DOE Partnership in Basic Plasma Science and Engineering, NASA's MMS mission 80NSSC18K0289, 80NSSC21K2048, and NSF Career Award 2142430. P.A.C. gratefully acknowledges support from NSF Grants AGS-1602769 and PHY-1804428, NASA Grants NNX16AG76G and 80NSSC19M0146 and DOE Grant DE-SC0020294. X.L.'s contribution is in part supported by NASA under grants 80NSSC18K0289 and 80NSSC21K1313. Simulations were performed with NERSC Advanced Supercomputing and LANL institutional computing. 

\section{Author contributions}
Y. L. performed and analyzed the simulations, derived the theory and wrote the paper; P. C. helped develop the theory and improved the writing; X. L. helped analyze the simulation data and develop the theory; M. H. helped develop the theory and interpreted the simulation data. S. L. provided inputs based on resistive MHD results. K. G. provided observational inputs and some insights. All authors discussed and commented on the paper. 

\section{Competing interests}
The authors declare no competing interests.\\

\newpage

\end{document}